\documentclass[11pt]{article}

\newcommand{\ignore}[1]{}%

\usepackage{algorithm, algorithmic}
\usepackage{amsmath}
\usepackage{amssymb}
\usepackage{amsthm}
\usepackage{tabularx}
\usepackage{environ}
\usepackage{graphicx}
\usepackage{subcaption}
\usepackage{url}
\usepackage[margin=1in]{geometry}
\usepackage{todonotes}

\usepackage{algorithm}
\usepackage{algorithmic}

\renewcommand{\algorithmiccomment}[1]{\bgroup\hfill\footnotesize~#1\egroup}

\usepackage[framemethod=tikz]{mdframed}
\mdfsetup{
	roundcorner=4pt,
	nobreak=true,
	linewidth=.5,
	innerleftmargin=25pt,
	innerrightmargin=25pt,
	innertopmargin=15pt,
	innerbottommargin=15pt,
	skipabove=12,skipbelow=8pt
}

\theoremstyle{plain}
\newtheorem{theorem}{Theorem}
\newtheorem*{theorem*}{Theorem}
\newtheorem{lemma}[theorem]{Lemma}
\newtheorem{claim}[theorem]{Claim}
\newtheorem{proposition}[theorem]{Proposition}
\newtheorem{observation}[theorem]{Observation}

\theoremstyle{definition}
\newtheorem{definition}{Definition}

\theoremstyle{remark}

\newcommand{\eps}{\varepsilon}

\newcommand{\cD}{\mathcal{D}}

\title{Optimal Single-Choice Prophet Inequalities from Samples}
\author{Aviad Rubinstein\thanks{Computer Science, Stanford University. \texttt{aviad@cs.stanford.edu}.} \and Jack Z. Wang\thanks{Computer Science, Cornell University. \texttt{jackzwang@cs.cornell.edu}.} \and S. Matthew Weinberg\thanks{Computer Science, Princeton University. \texttt{smweinberg@princeton.edu}. Supported by NSF CCF-1717899.}}
\date{}

\begin{document}

\maketitle
\thispagestyle{empty}
\begin{abstract}
We study the single-choice Prophet Inequality problem when the gambler is given access to samples. We show that the optimal competitive ratio of $1/2$ can be achieved with a single sample from each distribution. When the distributions are identical, we show that for any constant $\eps > 0$, $O(n)$ samples from the distribution suffice to achieve the optimal competitive ratio ($\approx 0.745$) within $(1+\eps)$, resolving an open problem of~\cite{CorreaDFS19}.
\end{abstract}
\newpage

\section{Introduction}\label{sec:intro}
Consider the classic single-choice Prophet Inequality problem. Offline, there are $n$ distributions $\cD_1,\ldots, \cD_n$ presented to a gambler. For $i=1$ to $n$, a random variable $X_i$ is drawn independently from $\cD_i$ and revealed online. The gambler must then decide immediately and irrevocably whether to accept $X_i$ (and achieve reward $X_i$, ending the game), or reject $X_i$ (continuing the game, but never revisiting $X_i$ again). The goal of the gambler is to devise a stopping rule which maximizes their expected reward. The performance of potential stopping rules is typically measured by their \emph{competitive ratio} in comparison to a prophet (who knows all $X_i$ in advance and achieves expected reward $\mathbb{E}[\max_i \{X_i\}]$). Typically, prophet inequalities are designed assuming that the distributions presented offline are fully known. This paper focuses on the setting where the gambler is instead presented with offline samples from the $\cD_i$, rather than complete knowledge.

In the classic setting, seminal work of Krengel and Sucheston provides a strategy guaranteeing a competitive ratio of $1/2$, which is the best possible~\cite{KrengelS78}.\footnote{To see that no better than $1/2$ is possible, consider an instance where $X_1$ is deterministically $1$, and $X_2$ is $1/\eps$ with probability $\eps$, and $0$ otherwise. The prophet achieves $2-\eps$ (taking $X_2$ when it is large, and $X_1$ otherwise), while the gambler achieves only $1$ (they must decide whether to take $X_1$ without knowing if $X_2$ is large).} Samuel-Cahn later proved that simply setting a threshold equal to the \emph{median} of $\max_i\{X_i\}$ (i.e. a value $T$ such that $\Pr[\max_i \{X_i\} > v] = 1/2$) also achieves the optimal competitive ratio of $1/2$~\cite{Samuel-Cahn84}, and it was later shown that a threshold of $\mathbb{E}[\max_i\{X_i\}]/2$ suffices as well~\cite{KleinbergW12}. These last two thresholds are remarkably simple, but certainly require a non-trivial number of samples to estimate well.

Our first result establishes that \emph{a single sample} from each $\cD_i$ suffices to achieve the optimal competitive ratio of $1/2$. The algorithm is also exceptionally simple: if $\tilde{X}_1,\ldots, \tilde{X}_n$ denote independent samples from $\cD_1,\ldots, \cD_n$, simply set $\max_i \{\tilde{X}_i\}$ as a threshold.

\begin{definition}[Single Sample Algorithm] Given as input $\tilde{X}_1,\ldots, \tilde{X}_n$, set a threshold $T=\max_i \{\tilde{X}_i\}$ and accept the first random variable exceeding $T$.
\end{definition}

\begin{theorem}\label{thm:noniid} The Single Sample Algorithm guarantees a competitive ratio of $1/2$.
\end{theorem}

A subsequent line of works considers the special case where each $\cD_i$ is identical (which we'll refer to as $\cD$). Here, work of Hill and Kertz provided the first improved competitive ratio (of $1-1/e$)~\cite{HillK82}, and this was recently improved to the optimal competitive ratio of $\alpha \approx 0.745$~\cite{CorreaFHOV17}. Our second result establishes that a \emph{linear number of samples} from $\cD$ suffices to achieve the optimal competitive ratio, up to $\eps$. Since our algorithm simply replaces the quantile-based thresholds of \cite{CorreaFHOV17} with samples, we call it Samples-CFHOV (the five authors of~\cite{CorreaFHOV17}). The algorithm and analysis are fairly simple and we provide a formal description in Section~\ref{sec:iid}.

\begin{theorem}\label{thm:iid} With $O(n/\eps^6)$ samples, Samples-CFHOV achieves a competitive ratio of $\alpha -O(\eps)$.
\end{theorem}

\subsection{Related Work}
Over the past decade, prophet inequalities have been studied from numerous angles within the TCS community~\cite{ChawlaHMS10, Alaei11, KleinbergW12,GobelHKSV14, EsfandiariHLM15,Rubinstein16, FeldmanSZ16, RubinsteinS17,DuettingFKL17, EhsaniHKS18, AzarCK18, AdamczykW18, CorreaSZ19,AnariNSS19, GravinW19, DuttingK19, ChawlaMT19}. All of these works assume explicit knowledge of the given distributions. The limited prior work most related to ours considers sample access to the underlying distributions. On this front, Azar et al. consider prophet inequalities subject to combinatorial constraints, and establish that limited samples suffice to obtain constant competitive ratios in many settings~\cite{AzarKW14}. In comparison to this work, our paper considers only \emph{optimal} competitive ratios, and the simple single-choice setting.

In the i.i.d. model (each value is drawn from the same $\cD$), a $(1-1/e)$-approximation was first shown in~\cite{HillK82}, and recent work achieved the same guarantee with $n-1$ samples from $\cD$~\cite{CorreaDFS19}. This ratio was later improved to $\approx 0.738$~\cite{AbolhassaniEEHKL17}, and then to $\alpha \approx 0.745$~\cite{CorreaFHOV17}, where $\alpha$ is the optimal achievable competitive ratio~\cite{HillK82,Kertz86}. The most related work in this sequence to ours is~\cite{CorreaDFS19}, who establish that a competitive ratio of $\alpha-\eps$ is achievable with $O(n^2)$ samples (for any constant $\eps$), and that $\Omega(n)$ samples are necessary. They also establish a formal barrier to achieving $\alpha-\eps$ with $o(n^2)$ samples. In comparison, we circumvent their barrier to achieve a competitive ratio of $\alpha-\eps$ with $O(n)$ samples, resolving one of their open problems.

\vspace{-2mm}
\noindent\paragraph{Roadmap.} In Section~\ref{sec:prelim}, we provide brief preliminaries. Section~\ref{sec:noniid} contains our $2$-approximation with a single sample. Section~\ref{sec:iid} contains our $(\alpha-\eps)$-approximation with linearly many samples.

\section{Preliminaries}\label{sec:prelim}
There are $n$ distributions, $\cD_1,\ldots, \cD_n$. Online, a gambler sees a random variable $X_i$ drawn from $\cD_i$ one at a time, and must immediately and irrevocably decide whether to accept (and get reward $X_i$) or reject (and see $X_{i+1}$, throwing away $X_i$ forever). Strategies for a gambler are often termed \emph{stopping rules}, and the competitive ratio of a stopping rule is the worst-case ratio (over all possible $n$, and $\cD_1,\ldots, \cD_n$) between its expected reward and $\mathbb{E}[\max_i \{X_i\}]$.

Our algorithms will not have knowledge of any $\cD_i$, but instead will have access to samples. Our algorithms will treat these samples as the only offline input, and decide whether to accept or reject an element based only on the value of that element and the samples.\footnote{In principle, sample-based algorithms might also consider previously viewed elements, but our algorithms don't.} Here, we will count the number of samples from each distribution as our sample complexity.

We will also consider the \emph{i.i.d. setting}, where each $\cD_i = \cD$. Here, we will count the total number of samples from $\cD$ as our sample complexity. In this setting, we will let $\alpha \approx 0.745$ denote the optimal competitive ratio for an algorithm with knowledge of $\cD$.
\vspace{-2mm}
\noindent\paragraph{Continuous vs. Discrete Random Variables.} All of our algorithm definitions are straight-forward for continuous distributions. For distributions with point masses, the following ``reduction'' to continuous is needed. Instead of thinking of $\cD$ as a single-variate distribution, we will (overload notation and) think of $\cD$ as a bivariate distribution with the first coordinate drawn from $\cD$, and the second ``tie-breaker'' coordinate drawn independently and uniformly from $[0,1]$. Then $(X_1,t_1) > (X_2,t_2)$ if either $X_1 > X_2$, or $X_1 = X_2$ and $t_1 > t_2$. Observe that because the tie-breaker coordinate is continuous, the probability of having $(X_1,t_1) = (X_2,t_2)$ for any two values during a run of any algorithm is zero. Therefore, if we define $F_{\cD}(X,t):=\Pr_{(Y,u)\leftarrow (\cD,U([0,1]))}[(Y,u)<(X,t)]$, we have that $F_{\cD}(X,t) < F_{\cD}(Y,u) \Leftrightarrow (X,t) < (Y,u)$. We will not explicitly reference this tie-breaker random variable in the definition of our algorithms, but simply refer to $X\leftarrow \cD$ as the pair $(X,t)$.

\vspace{-2mm}
\noindent\paragraph{Adversaries.} Prophet Inequalities are typically studied against an \emph{offline adversary}. That is, the adversary simply picks the distributions $\cD_1,\ldots, \cD_n$ (and their indices), which is all presented to the gambler offline. Some prophet inequalities hold against the stronger \emph{almighty adversary}, which selects the set of distributions $\{\cD_1,\ldots, \cD_n\}$ offline, then decides in which order to reveal the random variables $X_1,\ldots, X_n$ based on their realization. Note that previous competitive ratios of $1/2$ in the non-i.i.d. setting hold against an almighty adversary, and Theorem~\ref{thm:noniid} does as well. Previous competitive ratios of $\alpha$ in the i.i.d. setting hold against the offline adversary (and are impossible to achieve against the almighty adversary), so Theorem~\ref{thm:iid} holds against the offline adversary as well.

\section{The Non-I.I.D. Case: Optimal Ratio with a Single Sample from Each $\cD_i$}\label{sec:noniid}

The Single Sample Algorithm proceeds as follows. It takes as input $\tilde{X}_i$ drawn independently from each $\cD_i$, sets a threshold $T = \max_i \{\tilde{X}_i\}$, and accepts the first element exceeding $T$. Our goal in this section is to prove Theorem~\ref{thm:noniid} that this algorithm obtains $\frac{1}{2}$ the reward, in expectation, of the omniscient prophet that always selects the highest value.

Our analysis will use the principle of deferred decisions: instead of first drawing the samples $\tilde{X}$, and then revealing the actual draws $X$, we will jointly draw $2n$ samples $Y_1,\ldots, Y_n, Z_1,\ldots, Z_n$, and then for each $i$ randomly decide which of $\{Y_i,Z_i\}$ is equal to $\tilde{X}_i$ and which is equal to $X_i$. Formally, consider the following Deferred-Decisions procedure for drawing $X, \tilde{X}$:
\begin{enumerate}
\item Draw $Y_1,\ldots, Y_n$ and $Z_1,\ldots, Z_n$ independently each from $\cD_1,\ldots, \cD_n$.
\item For ease of notation later, for all $i$, relabel so that $Y_i > Z_i$.
\item Independently, flip $n$ fair coins. If coin $i$ is heads, set $X_i:=Y_i$ and $\tilde{X}_i:=Z_i$. Otherwise, set $X_i:=Z_i$ and $\tilde{X}_i:=Y_i$.
\end{enumerate}

\begin{observation}The output of the Deferred-Decisions procedure correctly generates $\tilde{X}_1,\ldots, \tilde{X}_n$ and $X_1,\ldots, X_n$ as independent draws from $\cD_1,\ldots, \cD_n$.
\end{observation}

Our analysis will proceed by directly comparing, for any fixed $Y_1,\ldots, Y_n, Z_1,\ldots, Z_n$, the expected reward of the gambler over the randomness in the coin flips of step three to the expected reward of the prophet over the randomness in the coin flips of step three. We note that this analysis is similar to that of the rehearsal algorithm of~\cite{AzarKW14} for $k$-uniform matroids (whose competitive ratio is asymptotically optimal for large $k$), and that prior to this it was folklore knowledge that the Single Sample Algorithm achieves a competitive ratio of at least $1/4$. The novelty in our analysis is precisely nailing down the tight competitive ratio.

\subsection{Analysis Setup}
For a fixed $Y_1,\ldots, Y_n, Z_1,\ldots, Z_n$, sort the values into descending order, and relabel them as $W_1,\ldots, W_{2n}$. If $W_j$ is equal to $Y_i$ (or $Z_i$), we say that $W_j$ \emph{comes from} $i$, and denote this with $\text{index}(W_j) = i$. Call the \emph{pivotal index}%
\footnote{The auhtors are grateful to Kaio Deeter for pointing out a mistake in the proceedings version of this definition.} $j^*$ the minimum $j$ such that there exists an $\ell < j$ with $\text{index}(W_\ell) = \text{index}(W_j)$. That is, the pivotal index $j^*$ is such that there are exactly $j^*-1$ $Y$ random variables exceeding the largest $Z$ random variable.

Our analysis will make use of the following concept: for each $W_1,\ldots, W_{j^*-1}$, let $C_j$ denote the outcome of the coinflip for $\text{index}(W_j)$ (which assigns either $Y_i$ or $Z_i$ to arrive as a sample and the other to arrive as a real value). Observe, importantly, that the random variables $C_1,\ldots, C_{j^*-1}$ are independent (because they are independent coin flips for different indices). Also importantly, observe that the random variable $C_{j^*}$ is deterministic conditioned on $C_1,\ldots, C_{j^*-1}$ (because it is exactly the same coin flip as one of the earlier indices).

\subsection{The Prophet's Expected Reward}

\begin{proposition}\label{prop:prophet} For fixed $W_1,\ldots, W_{2n}$ and pivotal index $j^*$, the prophet's expected reward, over the randomness in the coin flips of step three, is $\sum_{j=1}^{j^*-1} W_j/2^j + W_{j^*}/2^{j^*-1}$.
\end{proposition}
\begin{proof}
Observe that the prophet achieves expected reward equal to $\max_i\{X_i\}$, so we just want to compute the probability that this is $W_1,\ldots, W_{2n}$. For each $j < j^*$, $W_j$ is equal to $\max_i\{X_i\}$ if and only if $C_j$ is heads, and $C_\ell$ is tails for all $\ell < j$ (recall that all $W_j$'s are $Y$ random variables for $j < j^*$). Because each of the coin flips are independent, this occurs with probability precisely $1/2^j$.

For $j^*$, $W_{j^*}$ is equal to $\max_i\{X_i\}$ if and only if $C_\ell$ is tails for all $\ell < j^*$, and coin $C_{j^*}$ is tails. Observe, however, that by definition of the pivotal index $j^*$, that when $C_\ell$ is tails for all $\ell < j^*$ we have $C_{j^*}$ as tails as well (because it is the same coin as one of the first $j^*-1$). Therefore, whenever all of the first $j^*-1$ coins are tails, $\max_i\{X_i\} = W_{j^*}$ (and this happens with probability $1/2^{j^*-1}$).

As the first $j^*-1$ coins either contain some heads, or are all tails (and we have counted the prophet's reward in all such cases), we have now fully accounted for the prophet's expected reward over the randomness in the coin flips.
\end{proof}

\subsection{The Single Sample Algorithm's Expected Reward}
\begin{proposition}\label{prop:gambler} For fixed $W_1,\ldots, W_{2n}$ and pivotal index $j^*$, the gambler's expected reward, over the randomness in the coin flips of step three, is at least $\sum_{j=1}^{j^*-2} W_j/2^{j+1}+W_{j^*-1}/2^{j^*-1}$.
\end{proposition}
\begin{proof}
Consider the case where $C_1$ is tails. In this case, the gambler gets no reward because the threshold is higher than all revealed elements. For $j < j^*-1$, consider next the case where $C_1,\ldots, C_j$ are heads, but $C_{j+1}$ is tails. In this case, the gambler gets reward \emph{at least} $W_j$ (because the gambler will accept the first non-sample random variable exceeding $W_{j+1}$, and these random variables have values $W_1,\ldots, W_j$). The probability that this occurs is exactly $1/2^{j+1}$.

Consider also the case where $C_1,\ldots, C_{j^*-1}$ are all heads. Then the threshold is set at $W_{j^*}$, and the gambler will get at least $W_{j^*-1}$. This occurs with probability exactly $1/2^{j^*-1}$.
\end{proof}

\subsection{Proof of Theorem~\ref{thm:noniid}}
\begin{proof} We can immediately see that:
\begin{align*}
\sum_{j=1}^{j^*-2} W_j/2^{j+1} + W_{j^*-1}/2^{j^*-1} &\geq \sum_{j=1}^{j^*-1} W_j/2^{j+1} + W_{j^*}/2^{j^*}\\
&= \frac{1}{2}\cdot \left(\sum_{j=1}^{j^*-1} W_j/2^j + W_{j^*}/2^{j^* -1}\right).
\end{align*}
By Propositions~\ref{prop:prophet} and~\ref{prop:gambler}, the right-hand side is exactly half the prophet's expected reward, conditioned on $W_1,\ldots, W_{2n}$ and $j^*$, and the left-hand side is exactly the gambler's expected reward (again conditioned on $W_1,\ldots, W_{2n}$ and $j^*$). As the gambler achieves half the prophet's expected reward for all $W_1,\ldots, W_{2n}$ and $j^*$, the guarantee holds in expectation as well.
\end{proof}

\section{The I.I.D. Case: Optimal Ratio with Linear Samples from $\cD$}\label{sec:iid}
We begin with a brief overview of the algorithm from~\cite{CorreaFHOV17} and its main features, followed by a formal specification of our algorithm.

\subsection{Overview of~\cite{CorreaFHOV17} and Samples-CFHOV}
The algorithm of~\cite{CorreaFHOV17} (with one slight modification due to~\cite{CorreaDFS19}) proceeds as follows. We'll refer to this algorithm as Explicit-CFHOV.
\begin{enumerate}
\item As a function only of $n$, and independently of $\cD$, define monotone increasing probabilities $0 \leq p_1\leq \ldots p_n \leq 1$.
\item For all $i$ such that $p_i \leq \delta = \varepsilon^2/n$, update $p_i :=0$ (this is the~\cite{CorreaDFS19} modification).
\item Accept $X_i$ if and only if $F_{\cD}(X_i) > 1- p_i$. Observe that this is identical to accepting $X_i$ if and only if $X_i > \sigma_i = F_{\cD}^{-1}(1-p_i)$. Also observe that $X_i$ exceeds $\sigma_i$ with probability $p_i$.
\end{enumerate}

\begin{theorem}[\cite{CorreaFHOV17,CorreaDFS19}]\label{thm:explicit} In the i.i.d. setting, Explicit-CFHOV has competitive ratio $\alpha-\eps$.\footnote{Without step two, the algorithm achieves a competitive ratio of $\alpha$.}
\end{theorem}

That is, Explicit-CFHOV sets, for each $i \in [n]$, a probability $p_i$ independent of $\cD$, and sets a threshold $\sigma_i$ for accepting $X_i$ which is exceeded with probability exactly $p_i$.

If instead of explicit access to $\cD$, we're given $m$ i.i.d. samples from $\cD$, the challenge is simply that we can no longer compute $F_{\cD}(X_i)$ exactly and run Explicit-CFHOV. The algorithm of~\cite{CorreaDFS19} observes that $m = O(n^2)$ samples suffices to estimate the quantiles sufficiently well. Our algorithm observes that in fact $m= O(n)$ samples suffice (which is asymptotically tight, by a lower bound in~\cite{CorreaDFS19}). Our algorithm proceeds as follows, which we call Samples-CFHOV.

\begin{enumerate}
\item \textbf{As a function only of $n$, and independently of $\cD$}, define monotone increasing probabilities $0 \leq p_1\leq \ldots p_n \leq 1$, exactly as in Explicit-CFHOV.
\item Round down each $p_i$ to the nearest integer power of $(1+\eps)$; we denote the rounded value by $\lfloor p_i \rfloor \in \{(1+\eps)^{-1},(1+\eps)^{-2}\dots\}$.
\item Set $\tilde{p}_i := \lfloor p_i \rfloor/(1+\eps)$ (that is, we have rounded down each $p_i$, then further divided by $(1+\eps)$).
\item From our $m$ samples, let $\tau_i$ denote the value of the $(\tilde{p}_i\cdot m)$-th highest sample.
\item Accept $X_i$ if and only if $X_i > \tau_i$.
\end{enumerate}

That is, Samples-CFHOV provides an estimate $\tau_i$ of $\sigma_i$ via the $m$ samples. Intuitively, we are trying to overestimate $\sigma_i$ so that it is unlikely that Samples-CFHOV will ever choose to accept an element that Explicit-CFHOV would not. We'll prove Theorem~\ref{thm:iid} as a corollary of Theorem~\ref{thm:main}:

\begin{theorem}\label{thm:main} For any distribution $\cD$, with $m = O(n/\eps^6)$ samples, the expected value achieved by Samples-CFHOV is at least a $(1-O(\eps))$-fraction of that of Explicit-CFHOV.
\end{theorem}

We briefly remark that our proof of Theorem~\ref{thm:main} actually holds for any choice of $p_i$'s (all $\notin (0,\delta)$). That is, if Explicit-CFHOV achieves a competitive ratio of $\gamma(\vec{p})$ with a particular choice of $\vec{p}$, Samples-CFHOV achieves a competitive ratio of $\gamma(\vec{p}) - O(\eps)$ (as long as each $p_i \notin (0,\delta)$).

\subsection{Brief Comparison to~\cite{CorreaDFS19}}
The algorithm employed by~\cite{CorreaDFS19} using $O(n^2)$ samples is conceptually similar in that they also wish to set thresholds $\tau_i$ such that $F_{\cD}(\tau_i) \approx 1-p_i$. The main difference is that we target a multiplicative $(1-\varepsilon)$-approximation to each, whereas they target an additive $1/n$-approximation for each threshold. That is, they aim to ensure that for each $p_i$, the threshold $\tau_i$ has $|F_{\cD}(\tau_i) - p_i| \leq 1/n$. They prove, using the Dvoretzky-Kiefer-Wolfowitz Inequality, that $O(n^2)$ samples suffice for this, then further argue that these small additive errors in the CDF don't cost much.

The same paper also establishes a barrier to moving beyond $\Omega(n^2)$ samples. Specifically, they establish that $\Omega(n^2)$ samples are necessarily just to guarantee for a \emph{single} $i$ with $p_i \approx 1/2$ that $|F_{\cD}(\tau_i)-p_i| \leq 1/n$. Our approach circumvents this bound by seeking a significantly weaker guarantee for such $i$ (only that $|F_{\cD}(\tau_i) - p_i| \leq O(\varepsilon p_i)$ --- see Equation~\eqref{eq:concentration}). So the two key differences in our approach is (a) we show that $O(n)$ samples suffice to learn the thresholds up to a multiplicative $(1+\varepsilon)$ error in the CDF and (b) establishing that this (significantly weaker) estimation suffices for a good approximation.

\subsection{Proof of Theorem~\ref{thm:main}}
Our proof breaks down into two simple claims. The first establishes that with high probability, our sample-based thresholds are ``good.'' The second establishes that ``good'' thresholds yield a good approximation. Below, recall that $\delta = \varepsilon^2/n$.

\begin{lemma}\label{lem:main1}
With $m = 12\ln(1/\varepsilon)/(\varepsilon^3\delta) = O(n/\eps^6)$ samples, with probability at least $1-\varepsilon$, we have that for every $i$ (simultaneously),
\begin{gather}\label{eq:concentration}
 \frac{p_i}{(1+\eps)^3} \leq \Pr_{x \sim \cD}[x > \tau_i] \leq p_i.
\end{gather}
\end{lemma}

Note that Equation~\eqref{eq:concentration} does not reference the values of the actual elements $X_1,\ldots, X_n$ at all --- it is just a claim about the thresholds $\vec{\tau}$ being set. That is, the probability $1-\eps$ is taken only over the randomness in {\em drawing the samples} (and in particular independent of the values of the actual elements). We will call a set of thresholds ``good'' if they satisfy Equation~\eqref{eq:concentration}.
\begin{proof}
Recall that $\tau_i$ is set by first rounding down $p_i$ to $\lfloor p_i\rfloor$, then further dividing by $(1+\eps)$ to $\tilde{p}_i$, then set equal to the $(\tilde{p}_i \cdot m)$-th highest of $m$ samples. To proceed, let $L_i$ be such that
\[\Pr_{x \sim \cD}[x > L_i] = \lfloor p_i \rfloor.\]
Similarly, let $H_i$ be such that \[\Pr_{x \sim \cD}[x > H_i] = (1+\eps)^{-2} \lfloor p_i \rfloor.\]
Then~\eqref{eq:concentration} certainly holds whenever $L_i < \tau_i < H_i$. The remainder of the proof establishes that we are likely to have $L_i < \tau_i < H_i$ for all $i$.

Indeed, observe that we expect to see $\lfloor p_i \rfloor m$ samples greater  than $L_i$.
We say that $\lfloor p_i \rfloor$ is {\em bad} if the number of samples greater than $L_i$ is \emph{not} in the range $\Big[(1+\eps)^{-1}(\lfloor p_i \rfloor m),(1+\eps)(\lfloor p_i \rfloor m)\Big]$.
Note that whenever neither $\lfloor p_i \rfloor$ nor $(1+\eps)^{-2}\lfloor p_i \rfloor$ is bad, then we indeed have $L_i < \tau_i < H_i$.

Because the number of samples greater than $p$ is an average of $m$ independent $\{0,1\}$ random variables with expectation $p$, the multiplicative Chernoff bound implies that the the probability that a particular $p$ is bad is upper bounded by:
\[
\Pr[\text{$p$ is bad}] < e^{-\varepsilon^2 pm/3}.
\]
If all $p \in \{(1+\eps)^{-1},\ldots, \delta\}$ are not bad, then our desired claim holds. Taking union bound over this $(1+\eps)$-multiplicative net, we have that the probability that some $p \in \{(1+\eps)^{-1},(1+\eps)^{-2}\ldots, \delta\}$ is bad is bounded by:
\[
\sum_{i=0}^{O(\ln(1/\delta)/\varepsilon)} e^{-\varepsilon^2 (1-\varepsilon)^{-i} \delta m/3} \leq \sum_{i=0}^\infty  e^{-\varepsilon^2 (1-\varepsilon)^{-i} \delta m/3}\leq \sum_{i=0}^\infty  e^{-\varepsilon^3 i \delta m/6} \leq e^{-\varepsilon^3 \delta m /12}
\]

Above, the first term is simply a union bound over all $p$ in this net. The second inequality follows as $(1-\varepsilon)^{-i} \geq \varepsilon i/2$ for all $\varepsilon \in (0,1)$ and $i \geq 0$. The final inequality holds (at least) when $m \geq 6/(\varepsilon^3\delta)$. Therefore, setting $m = 12\ln(1/\varepsilon)/(\varepsilon^3\delta)$ satisfies the desired claim.
\end{proof}

Next, we wish to show that whenever the thresholds are ``good'', the algorithm performs well in expectation. Below, let $t_1$ denote the stopping time of Explicit-CFHOV (i.e. the random variable denoting the element it chooses to accept), and let $t_2$ denote the stopping time of Samples-CFHOV.

\begin{claim}\label{claim:stop}
Conditioned on~\eqref{eq:concentration} holding for every $i$, $t_2 \geq t_1$. That is, Samples-CFHOV selects an element \emph{later} than Explicit-CFHOV.
\end{claim}
\begin{proof}
For every $i$, we have that by~\eqref{eq:concentration}, the threshold used by Samples-CFHOV is at least the threshold used by Explicit-CFHOV. Therefore, the first time they deviate (if any) is when Explicit-CFHOV accepts an element but Samples-CFHOV does not.
\end{proof}

\begin{lemma}\label{lem:main2}
Conditioned on \eqref{eq:concentration} holding for every $i$, the following holds for every $v$:
\begin{gather}\label{eq:approx}
\Pr[X_{t_2} > v] \geq \frac{1}{(1+\eps)^3}\Pr[X_{t_1} > v].
\end{gather}
\end{lemma}
\begin{proof}
We prove that~\eqref{eq:approx} holds uniformly for every $i \in [n]$, i.e.
\begin{gather}\label{eq:approx-i}
\Pr[(X_{t_2} > v) \wedge (t_2 = i) ] \geq \frac{1}{(1+\eps)^3}\Pr[(X_{t_1} > v) \wedge (t_1 = i)].
\end{gather}
The event $(X_{t_b} > v) \wedge (t_b = i)$ (for either $b \in \{1,2\}$) occurs if and only if the corresponding algorithm \emph{doesn't} stop before $i$, and $X_i$ is larger than both $v$ and the threshold set (by the corresponding algorithm). Of course, whether or not an algorithm stops before $i$ is completely independent of $X_i$. We claim that the following holds on the probability that the two algorithms accept $X_i$ (conditioned on making it to $X_i$). Intuitively, Claim~\ref{claim:accept} establishes that, even though the threshold $\tau_i$ is stricter than $\sigma_i$, we are still roughly as likely to accept an $X_i$ exceeding $v$ using $\tau_i$ versus $\sigma_i$, for all $v$.

\begin{claim}\label{claim:accept} Conditioned on~\eqref{eq:concentration} holding for every $i$, then for every $v$ and $i$ such that $p_i \geq \delta$:
$$(1+\varepsilon)^3\Pr[(X_i > v) \wedge (X_i > \tau_i)] \geq \Pr[(X_i > v) \wedge (X_i > \sigma_i)].$$
\end{claim}
\begin{proof} We consider the following three cases: perhaps $v > \tau_i$, or perhaps $v \in (\sigma_i, \tau_i)$, or perhaps $v < \sigma_i$. We claim that the following three inequalities hold:
\begin{align*}
v \geq \tau_i &\Rightarrow \Pr[(X_i > v) \wedge (X_i > \tau_i)] = \Pr[X_i > v] = \Pr[(X_i > v) \wedge (X_i > \sigma_i)].\\
v \in (\sigma_i, \tau_i) &\Rightarrow \Pr[(X_i > v) \wedge (X_i > \sigma_i)] \leq \Pr[X_i > \sigma_i] \leq \frac{\Pr[X_i > \tau_i]}{(1+\varepsilon)^3} = \frac{\Pr[(X_i > v) \wedge (X_i > \tau_i)]}{(1+\varepsilon)^3}.\\
v < \sigma_i &\Rightarrow \Pr[(X_i > v) \wedge (X_i > \sigma_i)] = \Pr[X_i > \sigma_i] \leq \frac{\Pr[X_i > \tau_i]}{(1+\varepsilon)^3} = \frac{\Pr[(X_i > \tau_i )\wedge (X_i > v)]}{(1+\varepsilon)^3}.
\end{align*}

Indeed, the first implication follows as $v$ exceeds both $\sigma_i$ and $\tau_i$. The second implication follows as $v >\sigma_i$, and then by condition~\eqref{eq:concentration}. The third implication follows from condition~\eqref{eq:concentration}.
\end{proof}

Claim~\ref{claim:accept} is the heart of the proof, and we can now wrap up. Observe that $\Pr[(X_{t_2} > v) \wedge (t_2 = i)] = \Pr[t_2 \geq i] \cdot \Pr[(X_i > v) \wedge (X_i > \tau_i)]$. Similarly, $\Pr[(X_{t_1} > v) \wedge (t_1 = i)] = \Pr[t_1 \geq i] \cdot \Pr[(X_i > v) \wedge (X_i > \sigma_i)]$. By the work above, $(1+\varepsilon)^3\Pr[(X_i > v) \wedge (X_i > \tau_i)] \geq \Pr[(X_i > v) \wedge (X_i > \sigma_i)]$. By Claim~\ref{claim:stop}, $\Pr[t_2 \geq i] \geq \Pr[t_1 \geq i]$. Therefore, we've proven the desired claim for every $i \in [n]$. As $\Pr[X_{t_b} > v] = \sum_i \Pr[(X_{t_b}>v) \wedge (t_b = i)]$, this completes the proof of Lemma~\ref{lem:main2}.
\end{proof}

\begin{proof}[Proof of Theorem~\ref{thm:main}]
The proof of Theorem~\ref{thm:main} now follows as a direct corollary of Lemmas~\ref{lem:main1} and~\ref{lem:main2}. Lemma~\ref{lem:main2} asserts that whenever the thresholds are ``good'', Samples-CFHOV achieves at least a $1/(1+\varepsilon)^3$ fraction of the expected reward of Explicit-CFHOV (this is because the expected reward of Samples-CFHOV is simply $\int_0^\infty \Pr[X_{t_2} > v] dv \geq \int_0^\infty \Pr[X_{t_1} > v] dv/(1+\eps)^3$, and the expected reward of Explicit-CFHOV is precisely $\int_0^\infty \Pr[X_{t_2} > v] dv$). Lemma~\ref{lem:main1} asserts that the thresholds are good with probability at least $1-\varepsilon$. So together, Samples-CFHOV achieves at least a $\frac{1-\varepsilon}{(1+\varepsilon)^3}$ of the expected reward of Explicit-CFHOV.
\end{proof}

\bibliographystyle{alpha}
\bibliography{MasterBib}

\end{document}